# Analogue of the Mpemba effect in classical mechanics

## Alexei V. Finkelstein[*]

Institute of Protein Research, Russian Academy of Sciences, 142290 Pushchino, Moscow Region, Russia
*Correspondence: afinkel@vega.protres.ru

The name "Mpemba effect" was given to the finding that "If two systems are cooled, the water that starts hotter may freeze first", confirmed by numerous of observations. Now this paradoxical statement obtained a more general form "the state that is initially more distant from its equilibrium state comes to this state earlier". Though seemingly violating the fundamental laws of physical chemistry, this effect has been experimentally demonstrated for aqueous and various other systems, up to quantum dots. It was widely discussed in the *American Journal of Physics*, *Nature*, and other highly reputable journals, but the fundamental physical mechanism(s) underlying this effect remained elusive. Here I describe a simple mechanical system demonstrating the Mpemba effect ("the state that is initially more distant from its equilibrium state comes to this state earlier"), and show what physical mechanism causes this effect in this system.

The Mpemba & Osborne's statement that "If two systems are cooled, the water that starts hotter may freeze first" became known as the "Mpemba effect"[1] almost 60 years ago. Since then, this counterintuitive effect that seems to violate fundamental laws of physics has been observed in several systems: not only in freezing liquids like sweet milk, where this effect was – quite accidentally – observed for the first time[1], and water[2], but also in polymer crystallysation[3], clathrate hydrates[4], and quantum ferrimagnet systems[5], while numerical simulations have predicted a Mpemba-like behavior in spin glasses[6], granular fluids[7], and even quantum Ising models[8] and quantum dots[9].

The Mpemba effect has been reported in many experimental and even more numerous theoretical works published, among others, in the *American Journal of Physics, Proceedings of the National Academy of Sciences of the USA, Nature*, and *Nature Portfolio*. Despite the skepticism of some authors[10,11] regarding the reproducibility of the Mpemba effect and especially the homogeneity of the temperature across a liquid sample, this effect became famous because it was truly surprising that such an important and widespread phenomenon as water freezing had not yet been fully solved and understood.

However, the phenomenon now called the "Mpemba effect" (or "Mpemba paradox") was first mentioned almost 2,400 years ago by Aristotle, and much later, but still a very long time ago, by many famous natural philosophers, including Francis Bacon, René Descartes, Joseph Black, and others[12,13].

Since then, many explanations have been proposed for the Mpemba and Mpemba-like effects occurring in various systems. Most of them concern the faster cooling of hot liquid; here they are discussed in brief.

It goes without saying that some mineral salts dissolved in water may precipitate or leave the vessel walls during heating, or some dissolved gases may be released from the liquid when it boils[14,15], both changing the freezing temperature of the heated (or unheated) liquid. However, this concerns not the liquid itself, but rather methods of its purification and the state of the vessel walls.

Also, there is no doubt that evaporation or convection[14,16] can be stronger in hot liquids than in cold liquids. But here the following questions arise:

(i) Why do these phenomena persist during the entire cooling process of the "initially hot" liquid from its initial +80 °C to +20 °C (+20 °C is usually taken as the initial temperature of the "initially cold" liquid)?

(ii) Why are these phenomena still present in the "formerly hot" liquid (but not in the "initially cold" one) when both liquids are cooled from the same +20 °C to 0 °C – which takes many minutes?

(iii) What is the difference between the states of "initially hot" and "initially cold" liquids when both have a temperature of +20 °C?

The same questions apply to the "special construction" of hydrogen bond networks that appears to exist[17,18] in the "initially hot" liquid rather than in the "initially cold" one.

Despite extensive research, the fundamental physical mechanism(s) underlying the Mpemba and Mpemba-like effects remained elusive[19].

With the discovery of numerous Mpemba-like effects in various systems boosting the interest in such phenomena, the original Mpemba's statement "hot water freezes faster than cold water" has acquired the more general form of *"the state that is initially farthest from its equilibrium state attains the latter at the earliest time"*[5]. Still, the above questions remained unanswered.

## SIMPLE MECHANISTIC MODEL THAT EXHIBITS THE MPEMBA-LIKE EFFECT

To clarify the widely discussed Mpemba paradox, I will consider a simple mechanistic model that demonstrates the "counterintuitive" Mpemba effect, which is that *"the state initially farthest from its equilibrium attaining the latter at the earliest time."* The privilege of this mechanistic model is that it is completely described by the equations of motion.

Consider a point particle (red in Fig. 1A) that begins to falls, under the action of gravity, along a slope of the hill that is quite steep in its upper part and quite gentle at the foot, below the level $H$. The initial height of the particle is $h$, and its initial speed $v_h = 0$.

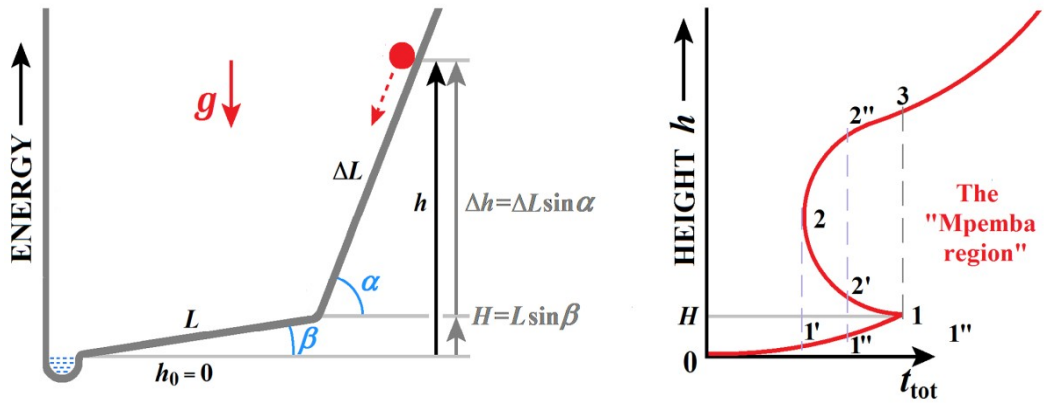

Fig. 1. (**A**) A point particle (red) falls from the hill under the action of gravity (the acceleration of gravity is $g$). The slope of the hill is quite steep in its upper part and gentle at the foot (the angle $\alpha$ is much larger than $\beta$). The particle begins to move at an initial height of $h$ with an initial speed $v_h = 0$. Simple question: How does the time $t_{tot}$ required for this particle to reach its most stable state (i.e. height $h_0 = 0$) depend on its initial height $h$? (**B**) Schematic representation of the dependence of the time $t_{tot}$ required for the point particle to reach its equilibrium position (final height $h_0 = 0$) on the initial height $h$ of this particle on the hill shown in Panel (**A**).

The introduced model in described by elementary Newtonian mechanics.

When the initial height $h$ of the point particle is below $H$ (Fig. 1), the particle reaches the level $h_0 = 0$ in time $t_{tot} = \frac{\sqrt{2h}}{\sqrt{g \cdot \sin\beta}}$, which only increases with its initial heights $h$ ($\frac{\partial}{\partial h} t_{tot} > 0$, see Fig. 1B, line 0–1'–1"–1); this is "normal", and no "Mpemba effect" can occur in this situation.

However, the situation is different when the initial position $h$ of the point particle is above $H$.

Obviously, now the particle reaches level $H$ in time $\frac{\sqrt{2\Delta h}}{\sqrt{g \cdot \sin\alpha}}$ with the speed $\sqrt{2g\Delta h}$, and then spends another time $\frac{\sqrt{2}L}{\sqrt{g\Delta h}+\sqrt{gh}}$ to descend to level $h_0 = 0$ (where it arrives with the speed $\sqrt{2gh}$). So, the total time of the descent is $t_{tot} = \sqrt{\frac{2}{g}}\left(\frac{\sqrt{h-H}}{\sin\alpha} + \frac{L}{\sqrt{h-H}+\sqrt{h}}\right)$. When $h$ grows, we have $\frac{\partial}{\partial h} t_{tot} = \frac{1}{\sin\alpha \cdot \sqrt{2g(h-H)}}\left(1 - \frac{L \cdot \sin\alpha}{\sqrt{h(h-H)}+h}\right)$. For $h > H$, $\frac{\partial}{\partial h} t_{tot}$ is negative when $h$ is only slightly larger than $H$, since $L = \frac{H}{\sin\beta}$ – see Fig. 1A, and $\alpha > \beta$, so that $L \cdot \sin\alpha > H$.

Thus, $\frac{\partial}{\partial h} t_{tot}$ changes its sign at the level of $H$: below $H$ it was positive, and just above $H$ it becomes negative (see the vicinity of point 1 in Fig. 1B).

The "strange" Mpemba effect is produced by this non-monotonic behavior of the $t_{tot}$ as a function of $h$. The value of $\frac{\partial}{\partial h} t_{tot}$ remains negative from the level $h=H$ to the level where $\frac{L \cdot \sin \alpha}{\sqrt{h(h-H)}+h}=1$, that is, where $h=H+L \cdot \frac{(\sin \alpha - \sin \beta)^2}{2 \sin \alpha - \sin \beta}$ (see the line 1–2'–2 in Fig. 1B); thus, the total descent time $t_{tot}$ decreases here. From this level of $h$, the total descent time $t_{tot}$ grows (see the line 2–2"–3 in Fig. 1B), and, when $h \gg H$, that is, when $h$ becomes big enough (after the point 3 in Fig. 1B), this total time can be approximated by a "normally looking", that is growing with $h$ dependence $t_{tot} \frac{\sqrt{2h}}{\sqrt{g} \cdot \sin \alpha}\left[1+\frac{H}{2h}\left(\frac{\sin \alpha}{\sin \beta}-1\right)\right]$.

Thus, the described system exhibits a "Mpemba-like" behavior when
(i) the initial height $h'$ of the "hotter" (excited) point particle is between points 1 and 3 in Fig. 1B;
(ii) the initial height $h$ of the "colder" point particle is below $h'$, *and*
(iii) this height $h$ is such that $t_{tot}(h)$ is greater than $t_{tot}(h')$ of the first particle.

If these three condition are satisfied, then this pair of particles exhibits the Mpemba effect: "*the state that is initially farthest from its equilibrium state attains the latter at the earliest time*".

Thus, the Mpemba effect may be "counterintuitive", but, being described by elementary equations of motion, it is by no means "strange".

Now we can answer all three questions concerning the behavior of this "Mpemba-like" system:
(i) Gravity persists all the time: it has the same accelerating effect on the particle independently of its current speed or position.
(ii) The inertia saves the kinetic energy obtained by the "formerly hot" particle while its descending to the level of the "initially cold" one, and the latter does not possess the kinetic energy at all at this level.
(iii) The difference between the states of the "initially hot" and "initially cold" particles when both are at the same height is the higher speed of the former.

**ACKNOWLEDGMENTS**

I am grateful to Bogdan S. Melnik and Andrei A. Klimov for discussions, and to E.V. Serebrova for editing the manuscript.